# Optimized Quantum Circuit Partitioning

Omid Daei[1], Keivan Navi[2] and Mariam Zomorodi-Moghadam[3]

**Abstract**
The main objective of this paper is to improve the communication costs in distributed quantum circuits. To this end, we present a method for generating distributed quantum circuits from monolithic quantum circuits in such a way that communication between partitions of a distributed quantum circuit is minimized. Thus, the communication between distributed components is performed at a lower cost. Compared to existing works, our approach can effectively map a quantum circuit into an appropriate number of distributed components. Since teleportation is usually the protocol used to connect components in a distributed quantum circuit, our approach ultimately reduces the number of teleportations. The results of applying our approach to the benchmark quantum circuits determine its effectiveness and show that partitioning is a necessary step in constructing distributed quantum circuit.

## 1. Introduction

In 1994, Peter Shor developed an algorithm for quantum processing which solves integer factorization problem by using quantum computing in polynomial time [1]. Shor's algorithm showed that quantum computers can solve that specific problem exponentially faster than the best known classical algorithm.

Quantum computers outperform their classical counterparts in solving certain problems such as search in database [2], discrete logarithm finding [1], and integer factorization [1] .

Quantum computing has many advantages over classical computing but realization of a quantum system on a very large scale is a serious challenge [3]. Experimental teams are developing powerful quantum processors to run small quantum algorithms [4]. But making an integrated quantum system with a large number of Qubits is too hard [5] [6]. Today's technology that implements quantum computing in the real world has limitations on the number of qubits able to be processed [8] and it is hard to make an integrated large-scale quantum computer [7] and this makes distributed implementation of quantum systems necessary [9].

Interaction of qubits with the environment is one of the limitations of implementing quantum systems [10]. While the number of qubits is increased, the quantum information becomes more sensitive to errors and the interaction of qubits results in decoherence [10].

Error correction codes also create a lot of overhead in the system, and a large number of qubits should be involved in

___________________________________________________________

Omid Daei
o.daei@qiau.ac.ir

Keivan Navi
navi@sbu.ac.ir

Mariam Zomorodi-Moghadam
m_zomorodi@um.ac.ir

[1] Department of Computer Engineering, Qazvin Branch, Islamic Azad University, Qazvin, Iran.

[2] Nanotechnology & Quantum Computing Lab and the Department of Computer Engineering and Science, Shahid Beheshti University, G. C., Tehran, Iran.

[3] Department of Computer Engineering, Ferdowsi University of Mashhad, Mashhad, Iran.

computing. Also, there is a possibility that it could not fit into a quantum chip [9].

Due to the above mentioned restrictions, one of the rational ways of making great quantum systems is to use distributed nodes where fewer qubits will be placed in each node or subsystem. So to build a large quantum computer, it is necessary to make a network of limited capacity quantum computers which are connected together through a classical or quantum channel and they all interact and simulate the behavior of a large quantum computer [11]. This scheme is known as distributed quantum computer.

For modeling distributed quantum circuits, we can use the model of monolithic quantum circuits. Each subsystem sends out data on demand to other ones through the communication channel created between the subsystems. In order to ensure that subsystems can communicate with each other in a distributed quantum system, there must be a reliable communication mechanism between the nodes of the distributed quantum system.

Realization of quantum communications can be regarded as a serious challenge [12]. One of the primitive communication protocols is teleportation [13] and some quantum technologies such as NMR [14] and trapped ions [15] have implemented teleportation [16] .

R. Van Meter [17] describes two different interconnect topologies known as teledata and telegate. In telegate approach a teleported gate is applied on the qubits without moving them. In teledata, without physically moving qubit states, information is teleported to another subsystem and is used to perform computations. In a VBE adder circuit, he has shown that the performance of teledata approach is better than of telegate by a factor of 3.5. Basically, since teleportation is a costly operation in distributed quantum systems, it is very important to reduce the number of teleportations in a distributed quantum system. On the other hand, based on no-cloning theorem, when a qubit is teleported to another node it cannot be used in the source node [18]. Regarding this issue, a method is proposed in this study to algorithmically distribute the qubits in nodes, in order to reduce the total number of teleportations needed.

The paper is organized as follows: in section 2, some of the basic concepts of quantum computing and distributed quantum computing are described. Related work is presented in section 3. Our proposed approach for designing a DQC with an improved number of teleportations is described in section 4. The results are specified in Section 5 and finally in Section 6, we conclude the paper with some suggestions for future work.

## 2. Background

In quantum computers, instead of bits, there are quantum bits or qubits. A bit can store zero or one, but a qubit can take zero, one, or any combination of zero and one at the same time which is called superposition. The basic states of a quantum system are displayed as $\{|0\rangle, |1\rangle\}$. Based on the quantum principles, a quantum system can not only be in its basic state, it can also be in a linear combination of basic states and the general state of a qubit is represented as $|\psi\rangle = \alpha|0\rangle + \beta|1\rangle$, where $|\psi\rangle$ is a quantum state and α and β are complex numbers so that $|\alpha|^2 + |\beta|^2 = 1$. In the classical circuits, gates act on bits and produce the desired output. Also in the quantum circuits, quantum gates have the same effect on qubits.

**Quantum gate:** A quantum gate is similar to a classic gate. When applied, a quantum gate changes the state of one or more qubits. An $n$-qubit quantum gate $U$ is defined by a $2^n \times 2^n$ matrix and by performing $U$ on a quantum state $|\psi\rangle$, the outcome is another state represented by the vector $U|\psi\rangle$.

Pauli gates, Hadamard, and Rotation are some commonly used single-qubit gates [16], [19]. If $U$ is a gate that operates on a single qubit, then controlled-$U$ is a gate which operates on two qubits and it works as follows: $U$ is applied to the target if the control qubit is $|1\rangle$ and leaves it unchanged otherwise. For example, Controlled-NOT (CNOT) is a two qubit gate. Target qubit changes if control qubit is $|1\rangle$ and unchanged otherwise. A quantum circuit (QC) consists of quantum gates interconnected by quantum wires. A quantum wire is a mechanism for moving quantum data from one location to another [20].

For the convenience of quantum computing, it is better to model them. There are several models for evaluating quantum computations, for example adiabatic model of computation [21] and quantum programming languages [22]. One of the main methods for displaying quantum computations is the circuit model, which is based on unitary evolution of qubits by networks of gates [16].

Figure 1 shows a simple quantum system in the circuit model which shows a quantum full adder.

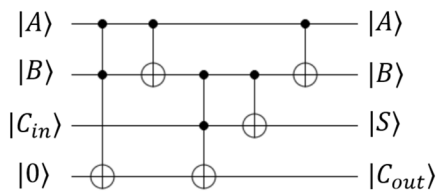

**Fig. 1** Circuit model of a quantum full adder

**Distributed quantum circuit:** A distributed quantum circuit (DQC) is an extension to the quantum circuit model. A DQC consists of *n* smaller QCs and each has a limited capacity and they are located far away from each other and all parts together implement the functionality of a QC. The different parts of a DQC should communicate with each other and send their qubits to each other by a quantum channel via teleportation.

**Local and global gates:** In a DQC system, a local gate is a gate whose all qubits reside in the same partition. On the other hand, a global gate is a gate whose qubits are in different partitions. . Therefore, to run this gate, the qubits information must be brought from the partitions in which they are located to the current partition where that gate runs.

Teleportation sends the state of a qubit from a point to another just by communicating two classical bits. Two points need to share a maximal entangled state [23]. QC for quantum teleportation is shown in figure 2 [16]. The first and second lines denote the sender's system and the third line is the receiver's system. Meters determine the measurement. Output of the measurement carries classical bits. Teleporting a quantum state works in any situation [13] [7], so it can be used as communication protocol for connecting partitions of a DQC to each other.

One of the advantages of using graph theory is to provide the same form for many problems. Formally, a graph is a pair of sets $(V, E)$, where $V$ is the set of vertices and $E$ is the set of edges, formed by pairs of vertices. In mathematics, a graph partitioning is the division of its nodes into mutually exclusive sets. The edges that connect the sets make the connection between those sets. Usually, graph partitioning problems are classified as NP-hard problems. So the solutions that have come up for such issues use heuristics and approximation algorithms, in general [24].

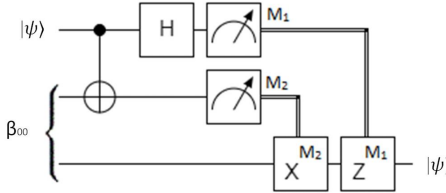

**Fig. 2** Quantum circuit for teleporting a qubit [16]

Suppose a graph $G = (V, E)$, where $V$ and $E$ represent a set of $n$ vertices and a set of edges respectively. A balanced partition problem divides $G$ into $k$ parts at the same size, So that the weight of the edges between separate components must be minimum [25]. Local and global algorithms are two methods for this purpose.

Two well-known local methods for graph partitioning are the Kernighan–Lin and Fiduccia-Mattheyses algorithms. They have local search for 2-way cuts. An effective drawback of these two algorithms that can affect the quality of the final result is the arbitrary initialization of the vertex set, but global approaches do not require an arbitrary initial partition and depend on the properties of the entire graph. Spectral partitioning is the most widely used method.

Multi-level graph partitioning algorithms perform partitioning using one or more stages. In each step, the size of the graph is reduced by collapsing the edges and vertices, the smaller graph is partitioned and then this part of the main graph is mapped back and refined [26]. In multi-level scheme, a wide range of partitioning methods can be used. In some problems, this method can lead to fast execution time and extremely desirable results. In this paper, Kernighan–Lin is extended to a multi-level scheme and this altered algorithm is used for QC partitioning.

*A. Kernighan-Lin graph partitioning algorithm*

The Kernighan–Lin (K-L) [27] algorithm is a heuristic algorithm for finding partitions of graphs. The input to the algorithm is an undirected weighted graph $G = (V, E, W)$; where $V$ is the set of vertices, $E$ is the set of edges, and $W$ is the set of weights assigned to edges. K-L algorithm partitions $V$ into two subsets A and B. Partitioning is done in a way that minimizes the sum of the weights of edges which are crossing from A to B. If the edges are without weights, we assume the weights of the edges as one. At first the vertices are randomly placed in partitions. Then we have to calculate two parameters: first, the cost of communication (weight of edges) of each node with the other nodes in its partition and then the cost of the communication of each node with other partitions. These two parameters are represented by $I\ internal\ cost$ and $external\ cost$. Suppose the difference between these two parameters is shown with $D$ $(D = E - I)$. The values of $D$ for all nodes in the circuit is calculated and the gain between the two nodes x and y, which are divided into two sections A and B, is calculated as $g_{xy} = D_x + D_y - 2C_{xy}$. Since we are calculating the cost of moving two nodes x and y, so the cost between them should not be calculated and will be reduced by $2C_{xy}$. Gain value will be computed for each of the two nodes and the largest number which is called $\hat{g}1$ and indicates the relationship between the two vertices is selected and the two vertices are locked and the two vertices of the set A and B are deleted. Now we update the values of $D$ and perform calculations for the remaining nodes. This action continues with the remainder of the vertices till two sets be empty. In each step we find $\hat{g}2$, $\hat{g}3$, and etc.

The parameter $k$ is the number of each stage of execution, $G_k = max \sum_{i=1}^{k} \hat{g}_i$ is calculated in every stage. The largest value specifies which two nodes must be exchanged. In the first run of the algorithm, the two said nodes will be exchanged, and for the next round, all locked nodes will be unlocked and the algorithm continues. This algorithm will continue while $G_k > 0$. Pseudo code for K-L is shown in algorithm1.

| Algorithm1 Pseudo code for Kernighan-Lin |
|---|
| 1 function Kernighan-Lin(G(V,E)): |
| 2  Distribute the nodes balances in sets a and b |
| 3   while ($G_k$ > 0){ |
| 4    for all a in A and b in B compute D values |
| 5    the lists gvect, avect, and bvect be empty |
| 6    for (n := 1 to \|V\|/2) |
| 7     find a from A and b from B, such that g = D[a] + D[b] - 2*c(a, b) be maximum |
| 8     remove a and b from further consideration in this pass |
| 9     add g to gvect, a to avect, and b to bvect |
| 10    update D values for the elements of A = A \ a and B = B \ b |
| 11   end for |
| 12   find k which maximizes $G_k$, the sum of gvect[1],...,gvect[k] |
| 13   if ($G_k$ > 0) then |
| 14 Exchange avect[1],avect[2],...,avect[k] with bvect[1],bvect[2],...,bvect[k] |
| 15  }//while |
| 16 return G(V,E) |

The K-L algorithm can be expanded to divide each partition into smaller partitions so that the communication between those partitions is optimal.

### B. Graph representation of a quantum circuit

Usually a weighted graph is defined as $G = (V, E, W)$. where $V$ is the set of vertices $V = \{v_1, v_2 ..., v_n\}$, $E$ is the set of edges $e_1, e_2, ..., e_k$ and $W$ is the set of edges weights represented as $w(e_1), w(e_2), ..., w(e_k)$. To transform a monolithic quantum circuit into a graph model, we must first represent the qubits and the connections between them in terms of vertices and edges in a graph model.

Suppose that $q_1, q_2, ..., q_n$ are qubits of QC, where $n$ is the number of qubits in the monolithic representation of the circuit. Representing qubits with vertices, then related graph has $n$ vertices and so the set $V$ of graph will be $V = \{q_1, q_2, ..., q_n\}$. Each gate in a QC is composed of one or more qubits. So qubits of QC are connected to each other through quantum gates. Edges of the graph can represent such relationship. If two qubits are related to each other through a gate, then there is an edge between their corresponding vertices in the graph, and otherwise the vertices are disjoint. Also the number of such considering a distributed quantum circuit, one-qubit quantum gates require only one qubit for execution, so they do not add any edges in the graph model. But for quantum gates with more than one qubit, their data must be brought to the point where the quantum gate is going to be executed. For example, suppose control and target of a CNOT gate are qubits $q_1$ and $q_2$. For performing this gate $q_1$ should be brought to the place where $q_2$ is located or vice versa. Therefore, a connection for this transfer is needed, which in the graph model we show it with an edge between $q_1$ and $q_2$ with initial weight of one. If both $q_1$ and $q_2$ are needed to run another gate, we increase the weight of the edge by one. This is repeated for all quantum gates to construct all the interconnections needed to run the gates in the graph model. Regarding all the interpretations, to convert a QC to a graph model, an undirected weighted graph is used. Figure 3 shows a sample circuit and its corresponding weighted graph.

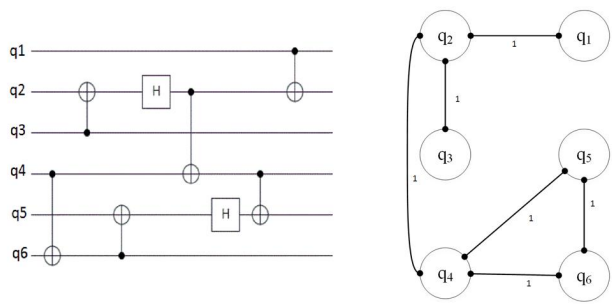

**Fig. 3** sample quantum circuit and corresponding weighted graph

## 3. Related Work

There are serious challenges to design and build devices that are not based on CMOS technology and are going to implement quantum computing [28]. Therefore to realize a quantum computer we will face very serious problems. One of this problems is due to the technology constraints. We cannot use a large number of qubits to build a single quantum device [29] and the greatest reason for the emergence of DQC is the existence of these constraints. Research on distributed quantum computing has been ongoing for nearly two decades [7]. Cleve and Buhrman [30] and Grover [31] were people who began to research on distributed quantum computing. After that, Cirac *et al.* [32] also carried out research in this field . In [31], Grover stated that to solve an issue in a distributed quantum system, the important question is how to split the problem into different quantum units in an optimal way. He divided a quantum system into different parts, each of which was separated and far from each other, and each section performed its computations separately. Every section that needed information from another part, received information from it. Grover showed that by the use of this distribution approach, the overall calculation time is reduced. He also introduced a quantum algorithm and matched it with distributed computation, although it possessed very costly communications.

In [30], Cleve and Buhrman have studied quantum communications and have shown that quantum entanglement can be used as a substitute for communication when we want to compute a function whose input data are distributed among remote locations. Also Cirac *et al.*, [32] showed that, for ideal quantum channels and for a sufficiently large number of nodes, the use of maximally entangled states is advantageous over uncorrelated ones.

Beals *et al.*, [33] introduced a distributed quantum system, each of parts was located on the nodes of a hypercube graph. They used a large number of qubits to apply a logical overhead to it. They showed that each QC can be converted into a DQC so that each part is at the vertex of a hypercube.

The author in [34] have presented an architecture with two communication methods for a distributed quantum computing which one of them used quantum communication between partitions and the other used classical communications. At the first type, each qubit can be entangled with any number of qubits and in the second type Small quantum systems are connected to each other by a network of classical communication channels.

The authors in [7] specified a language to illustrate QCs for distributed quantum computations. Also some definitions were presented for distributed quantum systems.

Connecting quantum computers via quantum internet to achieve faster processing speed is discussed in [35]. The challenges of designing quantum internet have also been studied in this work. The authors also discussed the creation of quantum internet in another article [36] and considered teleportation as the main strategy for the transmission of information. Then they explored the challenges and open issues in the design of quantum internet.

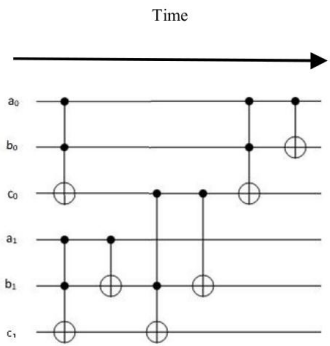

**Fig. 4** Details of a 2-qubit VBE adder in the monolithic form[17]

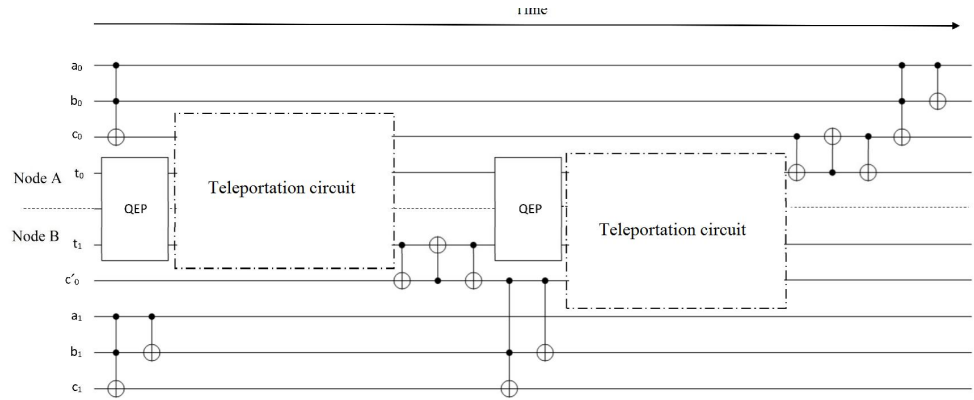

**Fig. 5** Details of a distributed 2-qubit VBE adder using the teledata method[17]

Hoi-Kwong Lo in [37] examined the cost of classical communication. The authors stated that, given the two-step teleportation, it takes a bit of classic communication at each step and $2\log_2 N$ bits of classical communications is needed for remote preparation of a desired N-dimensional state which is not the case in a normal teleportation.

The authors in [17] presented a fixed distributed quantum circuit of a monolithic quantum circuit. The circuit is a 2-qubit VBE adder and has been distributed in two nodes. It is divided into two equal parts and each part is located in a node and teleportation is used for the communication between these two nodes, illustrating in the figures 4 and 5. They calculated the cost of their teleportation design and compared two types of teleportations named teledata and telegate. Finally they concluded that teledata is better than telegate. We also used teledata for our distributed quantum cirucit teleportation. The authors in [28] compared their circuit with an integrated design and showed that when the node size rises, in large problems performance increases. Their architecture was fixed and they did not considered some issues related to DQC like teleportation cost reduction methods.

Especially they haven't made any effort to reduce the number of teleportations in their design which results in higher costs in large quantum circuits.

A distributed model of Shor's algorithm has been presented in [38]. Teleportation has been used as the communication protocol in this design. The design uses non-local gates for implementing the distributed quantum circuit for shor's algorithm. But no attempt has been made to determine the best location of qubits among distributed nodes and their method is not systematic like the one in [39]. Their method is fixed such that each register is divided into four parts and qubits of each part are hold in one computing node and output of each node is teleported to the next one. The authors did not try to optimize the number of teleportations. Also, they did not calculate the cost of returning the qubits back to their original part that were teleported.

In [40], the authors have examined the cost of quantum communication for DQCs by having just two partitions. They have computed the quantum cost of distributed components by providing an algorithm which searches for the best execution order of two-qubit CNOT global gates to reduce the total number of teleportations between the two partitions by examining all the possible ordering scenarios for performing global gates. And finally, the best execution order, which leads to the least amount of teleportation, have been explored. But the authors did not provide a solution for how to distribute qubits in different partitions to the minimum total cost needed and also their algorithm is limited to just two partitions and so it is not clear what problems are arisen when multiple partitions are considered. The authors improved their approach by using an evolutionary algorithm in [41]. Genetic algorithm has been used to find the best configuration for executing gates which are in two different partitions. The algorithm takes the teleportation cost as the cost function of the genetic algorithm and tries to assign the best partition for execution of each global gate.

The authors in [42] presented an automated way to distribute monolithic quantum circuits into multiple agents. They transform the input circuit into an equivalent one which consists of only Clifford+T gates and then swap the order between CNOTs and the 1-qubit gates from Clifford+T, pulling all CNOTs as early in the circuit as possible. This brings CNOTs closer together. But the authors have not considered the problem of monolithic quantum circuit partitioning with the purpose of reducing the number of communications. Also, they have not discussed the general arrangement of global gates in different partitions. they only discussed global gates that have special conditions.

**4. Proposed Approach**

In this section, we first describe the proposed method for the modeling of QCs to get a DQC ,and ultimately, at the end of this section, we describe the method in practice by applying it to a sample QC. Since teleportation in DQCs is a costly operation, it is worth reducing it as much as possible. Our approach is based on decreasing the number of teleportations in DQC.

First, we model the monolithic quantum circuit with an undirected weighted graph. Without loss of generality, it is assumed that the circuit is composed of one, two or three qubit gates.

Assuming a QC as input, it contains $n$ qubits $\{q_1, q_2, \ldots, q_n\}$, and $m$ gates,. Our gate library is a set $\mathcal{G}$ consists of one, two, and three qubit gates. The QC has a set $G = \{g_i \in \mathcal{G}; i: 1, \ldots, m\}$
($G = \{g_i(j, k, l) \text{ or } g_i(j, k) \text{ or } g_i(j) \in \mathcal{G}; i: 1, \ldots, m; j, k, l: 1, \ldots, n\}$)
of gates.

The number of qubits, $n$, in the QC determines the number of nodes in the graph model. So in the graph model, the set of vertices will be $V = \{q_1, q_2 \ldots q_n\}$.

To create an undirected weighted graph model, the connections between qubits in the gates are considered as the edges set $E$. Considering three different gate types, $E$ is defined as follows:
- For single-qubit gates: no edge is defined for them.
- For two-qubit gates:
$$E_i = \{(q_j, q_k) | g_i(j, k) \in QC; j, k: 1, \ldots, n\}$$
- For three-qubit gates: here we have three edges for each of the connections between qubits. We define them as follows:

$$E_{i1} = \{(q_j, q_k) | g_i(j, k, l) \in QC; j, k: 1, \ldots, n\}$$
$$E_{i2} = \{(q_j, q_l) | g_i(j, k, l) \in QC; j, k: 1, \ldots, n\}$$
$$E_{i3} = \{(q_k, q_l) | g_i(j, k, l) \in QC; j, k: 1, \ldots, n\}$$

As an example, if we have a two-qubit gate $g_i(1,2)$, which requires two qubits $q_1$ and $q_2$ to be executed, we represent the edge between $q_1$ and $q_2$ as $e_i = (q_1, q_2)$.

The edge set is constructed starting from the first gate in the QC. Once each edge is constructed, its weight is one. If there was more than one gate connecting two qubits in the QC, the weight of the corresponding graph edge is increased accordingly $w(e_1) = w(e_1) + 1$. Finally, we will obtain an undirected weighted graph which the weight of each edge, represents the number of gates requiring these two qubits for execution.

Since communications between partitions in a DQC are performed by teleportation, the number of these communications should be optimized. It is clear that we do not need to use teleportation for qubit interactions within a partition, because the qubits are available locally. Therefore, vertices must be arranged in different partitions in such a way that leads to the least number of communications between the partitions to run the DQC. Now, with a graph model of the QC, we can partition it into arbitrary parts using graph partitioning algorithms which considers the above issues. Given the implementation constraints and considering that the number of partitions affects the overall performance of the given method and this value varies for different circuits, we can calculate the values for the different partitions and finally select the desired partitions.

In order to have a balanced partitioned quantum circuit, each partition of the circuit must have almost the same number of qubits. Since the nodes in the graph represent qubits, the partitioning algorithm must be such that the nodes be distributed roughly equal in the partitions. To do this, different algorithms can be used. We used Kernighan-Lin (K-L) algorithm which is used in the VLSI design algorithms [27].

After applying the proper algorithm to graph model, we will have $K$ partitions $\{P1, P2 \ldots Pk\}$ in which there are a number of nodes and edges per partition and each partition may be connected to several other partitions through some edges.

Since graph partitioning algorithms attempt to partition a graph in order to optimize communications between partitions, so after applying the algorithm on the graph, it has the least communication outside the partition. This will result in the optimal number of teleportations required for the implementation of the DQC.

To construct the DQC out of the monolithic circuit, each part of the graph must be converted into an equivalent circuit. Each partition $P_i$ of the graph represents part of the DQC and the vertices in each partition represent qubits in that partition. The communication edges between them indicate a local gate between two qubits. Also, the most important part is the communication between the qubits that are not in a partition. The edges connecting the partitions represent global gates whose qubits are in more than one partition. It means, if there is an edge $e_{k1}$ from $q_i$ of partition $P_x$ to $q_j$ of partition $P_y$, and the weight of this edge shown with $w(e_{k1})$, So $w(e_{k1})$ teleportations between $q_i, q_j$ are needed. Sum of the weights of these edges among all partitions $W(E_K) = w(e_{k1}) + w(e_{k2}) + \cdots$ will eventually determine the number of teleportations which are required to run the DQC.

After the execution of the algorithm, the DQC will be obtained with improved communication which leads to the reduction in the number of teleportations.

The algorithm for the proposed approach is shown in Algorithm 2. First we build our graph-based model and then our partitioning algorithm which is based on K-L is applied on it.

a) Initially, $n$ (the number of vertices of related graph) is determined by the number of qubits and K is the number of arbitrary partitions.

b) The weight of all edges between vertices becomes zero.

c) For each gate in QC, if $q_i$ and $q_j$ are necessary to performing that gate, the weight of the edge between two vertices (qubits) increased one unit.

d) The graph is partitioned into desired number of partitions by calling the modified Kernighan-Lin function. So the vertices (qubits) in each partition are specified. Since K-L is an algorithm for binary partitioning, we modified K-L to use it recursively for the desired number of partitions.

f) Finally, all QC gates are placed in the DQC.

**Algorithm2** Algorithm for obtaining DQC from QC

Input: monolithic quantum circuit (QC), number of desired partitions (K),
Output: low cost distributed quantum circuit (DQC).

1: n= number of qubits of QC
2: **for** i ← 1 **to** n **do**
3:   **for** j ← 1 **to** n **do**
4:     Weight($e_{ij}$)=0
5:   **end for**
6: **end for**
7: **for** m ← 1 **to** last gate in QC **do**
8:   **if** gate(m) needs (qubit(i) and qubit(j)) to run **then**
9:     weight($e_{ij}$) = weight($e_{ij}$)+1
10: **end for**
11: Partition (1) **to** Partition (K) ← Kernighan-Lin (weight[1..n][1..n] , K)
12: **for** i ← 1 **to** K **do**
13:   place all nodes in partition(i) as a section
14: **end for**
15: **for** i ← 1 **to** last gate in QC **do**

16:  Add gate(i) in DQC
17: **end for**

The proposed method is explained by running on a sample QC shown in the figure 6. Each line represents a qubit. The connections between the qubits needed to run the gate are shown with a vertical line drawn between the qubits. After applying the proposed method on it, the corresponding graph model will be obtained and shown in figure 7.

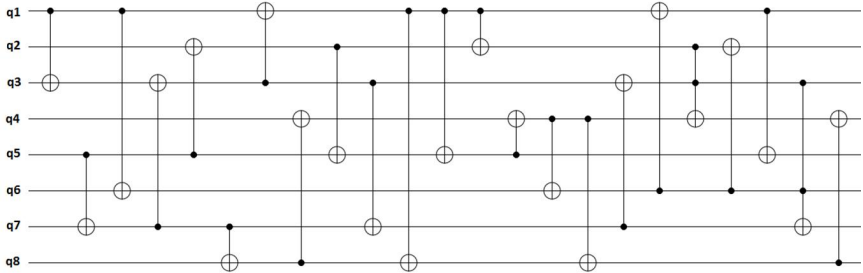

**Fig. 6** A sample QC

Because the QC has 8 qubits, we will have a graph with 8 vertices which are labeled one to eight, representing qubits from one to eight. Also, weight of the edges connecting the vertices shows the number of connections between those qubits on all the quantum gates of the sample QC. Now, with a graph model and applying the partitioning algorithms, it can be divided to the desired number of partitions.

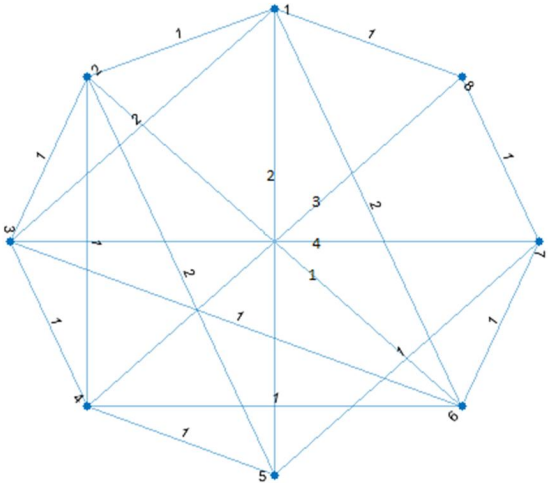

**Fig. 7** Graph corresponding to figure 6

For example, this graph is divided into 4 parts by K-L partitioning algorithm as follows P1{*q1,q6*}, P2{*q2,q5*}, P3{*q3,q7*}, P4{*q4,q8*}. Therefore, the QC is divided into four sections with specified qubits. The qubits that have the most connections are placed in a partition to reduce the number of teleportations that are needed. The DQC is shown in figure 8.

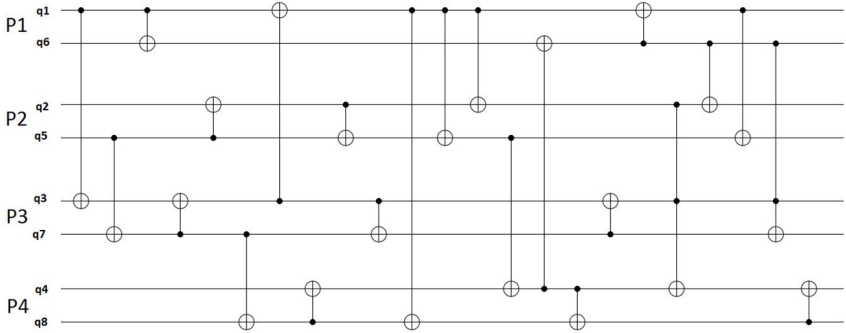

**Fig. 8** Sample QC split into 4 partitions

## 5. Results

We used MATLAB software to verify the proposed method experimentally. We also used some Revlib [43] library circuits as test circuits. Revlib is an online resource for benchmarking quantum and reversible quantum circuits. Another series of tests were conducted on the Quantum Fourier Transform with $n$ qubits (QFT for short) circuits where $n \in \{4, 8, 16, 32, 64\}$ and it is the number of qubits in the circuit. To implement and check the efficiency of the proposed method, the benchmark circuits must first be decomposed to basic gates library. The synthesis approach described in [44] and [45] was used for this purpose. Then we applied our proposed method to benchmark circuits for $k = 2, 3, 4$ number of partitions. As expressed, this partitioning is done in such a way that the least amount of communication between partitions is required and this will cause fewer global gates in distributed quantum circuits. In order to have a more accurate estimate, we implemented a random partitioning method for each of the given circuits. We performed this operation at 50,100 and 200 times, and finally we computed the average for each of them. This obtained average was the basis of comparison with our method. The results of the comparison of our approach and random partitioning in terms of number of global gates and improvement percentage for 12 different circuits are shown in table1.

**Table 1** Comparison of proposed approach with the random search (RS) of 50,100 and 200 repetition. The last column indicates percentage of decreasing global gates in the proposed approach in related to RS

| Circuit | # of qubits | # of partitions (K) | # of global_gates (P) | # of global_gates RS (#50) | # of global_gates RS(#100) | # of global_gates RS(#200) | % of improvement |
|---|---|---|---|---|---|---|---|
| QFT 4 BIT | 4 | 2 | 8 | 12 | 12 | 12 | 33% |
|  | 4 | 3 | 13 | 15 | 15 | 15 | 13% |
|  | 4 | 4 | 18 | 18 | 18 | 18 | 0% |
| QFT 8 BIT | 8 | 2 | 32 | 38 | 39 | 39 | 18% |
|  | 8 | 3 | 40 | 51 | 51 | 51 | 22% |
|  | 8 | 4 | 48 | 59 | 58 | 58 | 17% |
| QFT 16 BIT | 16 | 2 | 128 | 141 | 141 | 141 | 9% |
|  | 16 | 3 | 160 | 188 | 187 | 187 | 14% |
|  | 16 | 4 | 192 | 211 | 211 | 211 | 9% |
| QFT 32 BIT | 32 | 2 | 512 | 537 | 537 | 536 | 4% |
|  | 32 | 3 | 640 | 715 | 715 | 715 | 10% |
|  | 32 | 4 | 768 | 806 | 805 | 806 | 5% |
| QFT 64 BIT | 64 | 2 | 2048 | 2097 | 2097 | 2097 | 2% |
|  | 64 | 3 | 2560 | 2797 | 2796 | 2796 | 8% |
|  | 64 | 4 | 3072 | 3145 | 3145 | 3145 | 2% |
| FLIP_FLOP | 8 | 2 | 9 | 17 | 17 | 17 | 47% |
|  | 8 | 3 | 9 | 22 | 22 | 22 | 59% |
|  | 8 | 4 | 18 | 26 | 26 | 26 | 31% |
| FLIP_FLOP_NEW | 12 | 2 | 12 | 25 | 26 | 26 | 54% |
|  | 12 | 3 | 12 | 35 | 35 | 35 | 66% |
|  | 12 | 4 | 24 | 38 | 39 | 39 | 38% |
| ALU-PRIMITIVE | 6 | 2 | 8 | 13 | 13 | 13 | 38% |
|  | 6 | 3 | 10 | 17 | 17 | 16 | 38% |
|  | 6 | 4 | 13 | 18 | 18 | 18 | 28% |
| ALU1-28 | 5 | 2 | 5 | 8 | 8 | 8 | 38% |
|  | 5 | 3 | 5 | 11 | 11 | 11 | 55% |
|  | 5 | 4 | 8 | 12 | 12 | 12 | 33% |
| ADD16_174 | 49 | 2 | 4 | 97 | 99 | 99 | 96% |
|  | 49 | 3 | 8 | 132 | 131 | 130 | 94% |
|  | 49 | 4 | 12 | 147 | 147 | 147 | 92% |
| sym9_147 | 12 | 2 | 37 | 59 | 59 | 58 | 36% |
|  | 12 | 3 | 40 | 78 | 77 | 79 | 49% |
|  | 12 | 4 | 66 | 88 | 88 | 88 | 25% |
| rd32-272 | 5 | 2 | 8 | 11 | 11 | 11 | 27% |
|  | 5 | 3 | 8 | 14 | 14 | 14 | 43% |
|  | 5 | 4 | 13 | 15 | 15 | 15 | 13% |

In the qft (n) circuits, due to a particular pattern, almost all communications between qubits are equal and non-zero. With increasing the number of qubits, there is not much difference between the random partitioning and our approach. But in the circuits that do not have a uniform pattern or circuits with a uniform pattern and fewer connections between all qubits, there is a significant improvement in the number of global gates. Also, for qft (4) and k=4 partitions, there is no improvement. That is because the circuit has only 4 qubits, and with partitioning equal to 4, each partition has a qubit, so there is no difference between our approach and the random one.

In figures 9 and 10, Comparison the number of global gates for our approach (P) and random search by 200 repeated times RS (200) are shown for each of the 12 circuits of table 1. The results of qft circuits are shown in figure 9 and the other circuits are shown in figure 10. In the both the vertical axis represents the number of global gates and the horizontal axis shows the number of partitions which are K=2, 3 and 4.

Since the increase in the number of iterations of the random partitioning did not show any significant changes in the results, so we limited the iteration number to 200.

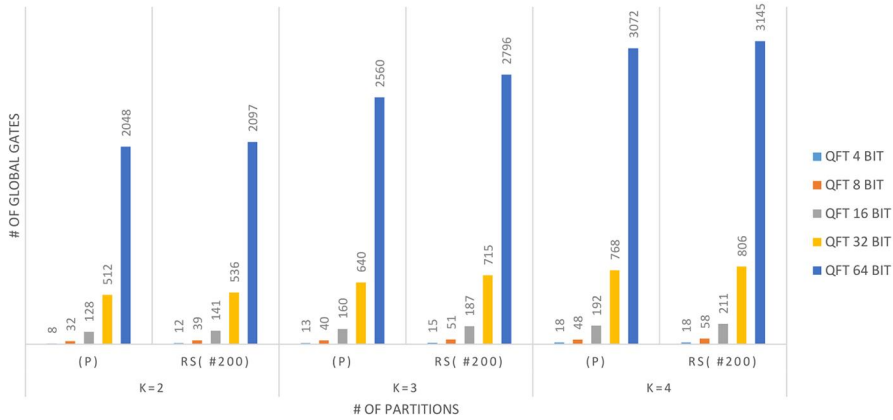

**Fig. 9** Number of global gates in (P) compared to RS in qft circuit

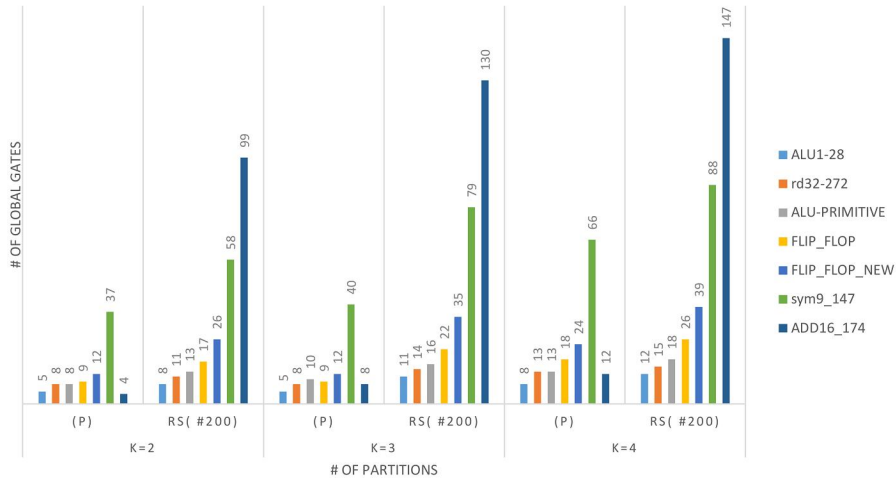

**Fig. 10** Number of global gates in (P) compared to RS(200) in seven other circuits

The percentage of improvement in proposed approach (P) relative to random search (RS) for each of the 12 benchmark circuits is shown in figure 11. The vertical axis represents the percentage of improvement in the number of global gates and the horizontal axis represents the number of partitions for each of the 12 circuits. Due to the use of our approach by reducing the number of global gates, the number of teleportations also decreases and the cost of teleportation in DQC will be reduced.

Using our method and with various number of partitions for each of the benchmark circuits, an average of 33% improvement has been achieved. It is necessary to note that since qft (4) is meaningless for 4 partitions, its results are not affected in this average. Also, if we calculate the percentage of improvement for benchmark circuits other than qfts, this value is increase significantly to an average of 47%.

**6. Conclusion and Future Works**

In this paper we presented a method for constructing distributed quantum circuits using a graph model of monolithic quantum circuits. Due to the fact that the teleportation in quantum technologies creates an overhead, the number of teleportations should be reduced as much as possible. Our approach to solving this issue is to model the circuit as a graph and to use a graph partitioning method in order to reduce the number of global gates in a DQC which ultimately reduces the number of required teleportations. Despite the similar existing study, our approach can be used to divide the quantum circuit into any desirable number of distributed systems. We compared our method with random partitioning method in order to ensure that our algorithm works well with the sample circuits.

After applying our proposed approach and partitioning the QC, each global gate in DQC can be executed in each of the partitions it belongs to. As future work, we aim to extend our algorithm to consider the location of global gates execution as part of the partitioning problem and to investigate its effect on the DQC cost.

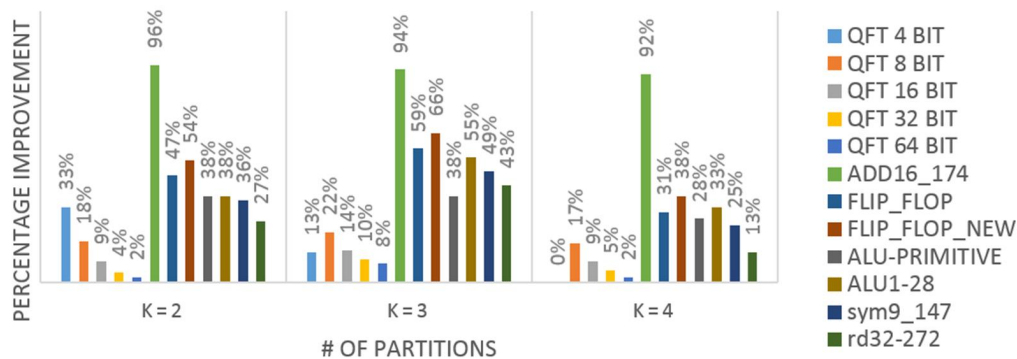

**Fig. 11** percentage of improvement comparing (P) with RS(200)